\documentclass[hyper,12pt,letterpaper]{JHEP3}
\usepackage{amsmath,amssymb,epsfig}

\font\cmss=cmss12

\newcommand{\bc}{\begin{center}}
\newcommand{\ec}{\end{center}}
\newcommand{\ba}{\begin{array}}
\newcommand{\ea}{\end{array}}
\newcommand{\beq}{\begin{equation}}
\newcommand{\eeq}{\end{equation}}
\newcommand{\bea}{\begin{eqnarray}}
\newcommand{\eea}{\end{eqnarray}}
\newcommand{\bmx}{\begin{pmatrix}}
\newcommand{\emx}{\end{pmatrix}}
\newcommand{\nn}{\nonumber}

\newcommand{\m}{\mu}

\newcommand{\Om}{\Omega}

\newcommand{\half}{\frac{1}{2}}
\newcommand{\tr}{{\rm tr}}

\newcommand{\eref}[1]{Eq.~(\ref{#1})}

\newcommand{\cF}{{\cal F}}

\newcommand{\ybar}{{\bar y}}

\newcommand{\tA}{{\tilde A}}
\newcommand{\tq}{{\tilde q}}
\newcommand{\tF}{{\tilde F}}

\newcommand{\hq}{{\hat q}}
\newcommand{\vol}{{\rm vol}}

\newcommand{\doublet}[2]{\genfrac{}{}{0pt}{}{#1}{#2}}
\def\IB{\relax{\rm I\kern-.18em B}}
\def\IC{{\relax\hbox{\kern.3em{\cmss I}$\kern-.4em{\rm C}$}}}
\def\ID{\relax{\rm I\kern-.18em D}}
\def\IE{\relax{\rm I\kern-.18em E}}
\def\IF{\relax{\rm I\kern-.18em F}}
\def\II{\relax{\rm I\kern-.18em I}}
\def\IZ{\relax{\sf Z\kern-.35em Z}}
\def\Id{\relax{1\kern-.32em 1}}
\def\IG{\relax\hbox{$\inbar\kern-.3em{\rm G}$}}
\def\IR{\relax{\rm I\kern-.18em R}}

\normalsize

\title{Noncritical-Topological Correspondence: Disc Amplitudes and
  Noncompact Branes} \author{Anindya Mukherjee\,\footnote{Email:
    anindya\_m@theory.tifr.res.in}\,, Sunil Mukhi\,\footnote{Email:
    mukhi@tifr.res.in}\,, and Rahul Nigam\,\footnote{Email:
    rahulnig@theory.tifr.res.in}\\ \it Tata Institute of Fundamental
  Research,\\ \it Homi Bhabha Rd, Mumbai 400 005, India} \abstract{We
  examine the duality between type 0 noncritical strings and
  topological B-model strings, with special emphasis on the flux
  dependence. The former theory is known to exhibit holomorphic
  factorisation upto a subtle flux-dependent disc term. We give a
  precise definition of the B-model dual and propose that it includes
  both compact and noncompact B-branes. The former give the factorised
  part of the free energy, while the latter violate holomorphic
  factorisation and contribute the desired disc term.  These observations
  are generalised to rational radii, for which we derive a
  nonperturbatively exact result. We also show that our picture extends to a
  proposed alternative topological-anti-topological picture of the
  correspondence for type 0 strings.}

\preprint{hep-th/0612054\\ TIFR/TH/06-38}


\begin{document}

\section{Introduction}
\label{Introduction}

It has been known for some time\cite{Witten:1991mk,Mukhi:1993zb,
  Ghoshal:1993qt,Hanany:1994fi,Ghoshal:1994rs} that noncritical string
theories in two spacetime dimensions have a topological
description. Subsequently the actual correspondence between them and
their topological duals on a Calabi-Yau manifold was found by Ghoshal
and Vafa\cite{Ghoshal:1995wm}. Noncritical $c=1$ string theory at the
self-dual radius is perturbatively equivalent to topological string
theory on a deformed conifold. For integer multiples of the self-dual
radius, the corresponding topological theory lives on a $\IZ_n$
orbifold of the conifold geometry. The noncritical-topological
equivalence was shown both via Landau-Ginsburg models and using the
ground ring\cite{Witten:1991zd} construction.

The above correspondence has led to considerable illumination of the
existence and properties of noncritical strings. But because the bosonic
$c=1$ string is nonperturbatively unstable, it has not been possible
to extend the equivalence to the nonperturbative level.  Such an
extension can be explored in the case of the nonperturbatively stable
Type 0 string theories in two dimensions. These too have a description
in terms of topological string theory, first proposed in
Ref.\cite{Ita:2004yn} and further explored in 
Refs.\cite{Danielsson:2004ti,Hyun:2005fq,Danielsson:2005uz}.
The Calabi-Yau dual to noncritical type 0 strings at a special radius
is a $\IZ_2$ orbifold of the conifold, while for integer multiples of
this radius it is a $\IZ_{2n}$ orbifold.

One of the most interesting aspects of noncritical type-0 strings is
the possibility of turning on background RR fluxes: type 0A theory has
two RR gauge fields and type 0B theory has an RR scalar whose
equations of motion admit linear growth in space and
time\cite{Douglas:2003up}. Thus in both cases there is a pair of
independent RR fluxes $q$ and $\tq$. At the level of the closed string
perturbation expansion the theory depends only on $|q|+|\tq|$, but at
the disc level there is a subtle and important additional term in the
free energy which depends on $|q|-|\tq|$, as found by Maldacena and
Seiberg in \cite{Maldacena:2005he}. Only after this extra term is
included in the free energy, one finds a satisfactory physical
interpretation wherein one of the two fluxes is sourced by ZZ 0-branes
while the other has no sources.

However, the dual topological B-model of Refs.\cite{Ita:2004yn} and
\cite{Hyun:2005fq} depends on the complex-structure moduli of the
orbifolded conifold, which in turn depend on the fluxes only through
the combination $|q|+|\tq|$.  Thus it is not obvious how the
correspondence can be extended to incorporate this effect.  Our aim
here will be to re-examine the duality with particular reference to
flux dependence. We will make the existing proposal more precise and
will then argue that the topological side must be
extended to incorporate a new feature, namely noncompact
branes\footnote{A tentative indication of a role for noncompact
  branes in this duality was found in Ref.\cite{Ita:2004yn}.}. When placed at
  appropriate locations, they contribute precisely the desired disc
  term in the free energy. Thereafter we generalise these observations
  to integer multiples of the special radius and to infinite
  radius. 

One key difference between our analysis and previous ones is that we
use the result of Ref.\cite{Maldacena:2005he} which we consider to be
rigorously true as a convergent integral representation of the full
(nonperturbative) free energy of type 0 strings.

\section{Noncritical-topological duality}
\label{Dualities}

\subsection{Bosonic case}

In this section we briefly review some relevant aspects of topological
string theory on noncompact Calabi-Yau spaces. The simplest example is
the conifold, described by the equation
\beq
\label{conifold}
zw-px = 0,
\eeq
where $z,w,p,x$ are complex coordinates of $\IC^4$. This is therefore
a three complex dimensional non-compact manifold. It has a singularity
at the origin. The singularity can be removed by blowing up an $S^3$
cycle at the origin, after which  the equation becomes:
\beq
\label{dc}
zw-px = \mu,
\eeq
where $\mu$ is in general complex and its modulus determines the size
of the $S^3$. \eref{dc} is known as the deformed conifold (DC) and
$\mu$ is its complex structure parameter. The singularity in
\eref{conifold} can alternatively be removed by blowing up an $S^2$ at
the origin. The resulting manifold is the resolved conifold (RC).  

The topological A model on any given Calabi-Yau is a theory of
quantised deformations of the Calabi-Yau, sensitive only to the
K\"ahler moduli. The B model is similar but depends only on the
complex structure moduli. The noncritical-topological duality proposed
by Ghoshal and Vafa\cite{Ghoshal:1995wm} stems from the
observation\cite{Witten:1991zd} that the ground ring of $c=1$ string
theory at the self-dual radius has four generators $z,w,p,x$ that are
worldsheet operators of conformal dimension 0 in the BRST cohomology,
satisfying the conifold relation \eref{dc} where
\beq
\mu=ig_s\mu_M
\eeq
and $\mu_M$ is the cosmological constant on the worldsheet in the
noncritical string theory\footnote{The factor of $i$ exhibited here is
  often dropped in the literature, though it has been correctly
placed in Refs.\cite{Danielsson:2004ti,Danielsson:2005uz}.  It
  is important because the genus expansion of the topological string
  free energy is $F^{top} = \sum_g \chi_g \mu^{2-2g}$, with
  coefficients that alternate in sign (given by the virtual Euler
  characteristic of the moduli space of genus $g$ Riemann surfaces),
  while that of the $c=1$ string is $F_{c=1}=\sum_g |\chi_g|
  (g_s\mu_M)^{2-2g}$ and is therefore positive in every genus as
  befits a unitary theory. A discussion of this point may be found in
  Ref.\cite{Mukhi:2003sz}.}.  Based on this and other evidence, they
argued that the $c=1$ string with cosmological constant $\mu_M$ at
self-dual radius is equivalent to the topological B-model on the
deformed conifold with deformation parameter $\mu$. In particular
their argument requires the genus-$g$ partition functions of the two
theories to coincide. Writing the genus expansions of the free
energies of the $c=1$ theory and the topological theory on the
deformed conifold as:
\bea
\cF^{c=1}(\mu_M) &=& \sum_{g=0}^\infty \cF_g^{c=1} \mu_M^{2-2g}\nn\\
\cF^{top,DC}(\mu)&=& \sum_{g=0}^\infty \cF_g^{top,DC} \mu^{2-2g}
\eea
the claim then amounts to:
\beq
\cF_g^{c=1} = (ig_s)^{2-2g}\cF_g^{top,DC},\quad {\rm all}~ g
\eeq
for which ample evidence has been
found\cite{Antoniadis:1995zn,Morales:1997zv}. There is also expected
to be a 1-1 correspondence between the physical observables (tachyons
in the $c=1$ case and deformations of $S^3$ in the B-model case) and
their correlators (for a recent discussion, see Ref.\cite{Gukov:2005iy}).

Going beyond the self-dual radius, it has long been
known\cite{Ghoshal:1992kx} that the ground ring of the $c=1$ string at
integer multiples of the self-dual radius, $R=p$, is a $\IZ_p$
orbifold of the conifold. This space has $p$ singularities connected
by (complex) lines. The deformed version of this space is described by
the equation:
\beq
\label{orbconif}
zw - \prod_{k=1}^p (px-\mu_k) =0
\eeq
which has $n$ homology 3-spheres of size $\mu_1,\mu_2,\ldots,\mu_p$,
each concealing one of the singularities. The geometry develops a
conifold singularity if any of the $\m_i$'s become zero, and a line
singularity if $\m_i=\m_j$ for $i\ne j$.  If the $\m_i$'s are all
distinct and nonzero, the manifold is non-singular.

We expect the $n$ deformation parameters to be in correspondence with
$p$ distinct (non-normalisable) deformations of the noncritical string
theory\cite{Ghoshal:1993ur}. If we only choose to perform the
cosmological constant deformation $\mu_M$ then these $p$ deformation
parameters must be determined in terms of $\mu_M$. It has been shown
via a Schwinger computation\cite{Gopakumar:1998vy} that for an integer
radius the parameters $\mu_k$ are given by $ig_s\frac{\mu_M +ik}{p}$,
and $k=-\frac{p-1}2, -\frac{p-1}2 +1, \cdots, \frac{p-1}2$. Moreover,
the free energy factorises\footnote{We use the word ``factorise'' even
  though the free energy splits into a sum, rather than a product, of
  terms. What factorises is of course the partition function.}
into a sum of contributions as follows:
\beq
\label{factformula} 
\cF_{c=1}^{R=p}(\m) = \cF^{top,DOC_p}(\{\m_k\}) =
\sum_{k=-\frac{p-1}2}^{\frac{p-1}2} \cF^{top,DC}(\m_k) 
\eeq

This factorisation can be understood in the Riemann surface
formulation of \cite{Aganagic:2003qj}. In this approach one thinks of
the following class of noncompact Calabi-Yaus:
\beq
zw-H(p,x)=0
\eeq
as a fibration described by the pair of equations:
\beq
zw=H,\quad H(p,x)=H
\eeq
The fibre is $zw=H$, a complex hyperbola, and the base is the
complexified $p,x$ plane. Above points in the base satisfying
$H(p,x)=0$, the fibre degenerates to $zw=0$, a pair of complex planes
intersecting at the origin. Such points in the base form a Riemann
surface, and it is this surface that governs the physics of the
topological string theory. Moreover the function $H(p,x)$ plays the
role of a Hamiltonian and lends an integrable structure to the system.

In the present case of the orbifolded conifold of \eref{orbconif}, the
Hamiltonian is:
\beq
H(p,x)=\prod_{k=1}^p (px-\mu_k)
\eeq
and hence the Riemann surface $H(p,x)=0$ factorises into disjoint
Riemann surfaces\cite{Hyun:2005fq}. This is the physical reason for
the factorisation of the free energy into a sum of contributions, one
for each branch of the Riemann surface.

The above statements are meaningful only at the level of string
perturbation theory, since the bosonic $c=1$ is not well-defined
nonperturbatively. Moreover, the computation of
Ref.\cite{Gopakumar:1998vy} is performed by manipulating a divergent
series. Later we will discuss the analogous relation for the type 0A
string, and will demonstrate factorisation of the free energy without
ever using perturbation expansions or divergent series. In this way we
will reliably show that it is nonperturbatively exact.

\subsection{Type 0 case, $R=1$}

In Ref.\cite{Ita:2004yn} and subsequently Ref.\cite{Danielsson:2004ti,
  Hyun:2005fq,Danielsson:2005uz}, the above ideas were applied to the
case of the type 0A string. Here it is convenient to choose units in
which $\alpha'=2$. A new feature of the type 0A string relative to the
bosonic case (for more details, see
Refs.\cite{Douglas:2003up,Gukov:2003yp,Maldacena:2005he} and
references therein) is that it has two distinct quantised parameters
$q$ and $\tq$. In the Liouville description these arise as the fluxes
of two distinct Ramond-Ramond 2-form field strengths,
$F_{t\phi},\tF_{t\phi}$. The theory has a symmetry, labelled
S-duality, under which the cosmological constant $\mu_M$ changes sign
and at the same time, $F\leftrightarrow \tF$. In the more powerful
matrix quantum mechanics (MQM) description of the same string theory,
the fluxes have quite an asymmetric origin. For $\mu_M<0$, $q$ is the
difference in the number of $D0$ and ${\bar D0}$ branes, or the net
number of $D0$ branes, in the MQM. On the other hand, $\tq$ is the
coefficient of a Chern-Simons term involving gauge fields on the
branes and anti-branes. For $\mu_M>0$ the roles of $q,\tq$ are
reversed. On reducing to eigenvalues, each of the integers $q$ and
$\tq$ can be interpreted as the quantised angular momentum of fermions
moving in the complex plane. Moreover, if both are turned on there is
an additional coupling term arising from projection to nonsinglet
sectors, such that the Hamiltonian eventually depends only on
$(q+\tq)$.

The Euclidean type 0A theory has a special value of the radius, $R=1$
in these units, at which the correspondence with the topological
string is simplest. This radius is the analogue of the self-dual
radius for the bosonic $c=1$ string.  For type 0A noncritical strings
at the special radius, the corresponding dual geometry in the
topological string has been proposed\cite{Ita:2004yn} to be a deformed
$\IZ_2$ orbifold of the conifold (DOC). The identification is again
based on the analysis of the ground ring of the noncritical theory.
The DOC dual to the type 0A string has two $S^3$'s whose complex
structure parameters are identified\footnote{Again, this identification
  substantially agrees with that in
  Refs.\cite{Danielsson:2004ti,Danielsson:2005uz} but differs from
  that in Refs.\cite{Ita:2004yn,Hyun:2005fq} by factors of $i$.}
 with the type 0A parameters $\mu_M, \hq=q+\tq$ as:
\bea\nn
\label{mudefs}
\mu =& ig_s(\mu_M -\frac{i\hq}{2}) &= \frac{g_s}{2}y\nn\\
\mu' =& -ig_s(\mu_M +\frac{i\hq}{2})&= \frac{g_s}{2}\ybar
\eea
with:
\beq
\label{ydef}
y=\hq+2i\m_M
\eeq
Thus the equation of the DOC is:
\beq
\label{doconif}
zw + (px-\mu)(px-\mu')=0
\eeq
Notice that complex conjugation exchanges the moduli of the two
$S^3$'s and acts as S-duality of the noncritical string. This is
because both conjugation and S-duality act as
$\hq\to\hq,\mu_M\to-\mu_M$. As a result the S-duality of type 0A
noncritical strings is explicitly geometrised in the topological
B-model dual. 

We note at this point that a different point of view about
noncritical-topological duality for type 0 strings is espoused in
Ref.\cite{Danielsson:2005uz}, according to which the topological
string is defined on the ``holomorphic square root'' of the space we
have been discussing, which is an ordinary conifold rather than an
orbifolded one. The noncritical-topological correpondence then has to
be reformulated by saying that we have to add topological and
anti-topological free energies. While this seems to fit in with the
picture of topological strings emerging from black hole
studies\cite{Ooguri:2004zv,Ooguri:2005vr}, it is not clear that in
practical terms it differs from the older proposal of
Ref.\cite{Ita:2004yn}. However we will see later that our proposal for a
precise topological dual involving noncompact branes can also be
phrased in topological-anti-topological language.

The manifold \eref{doconif} exists and is nonsingular for all
nonzero $\mu\ne\mu'$. However, the topological B-model on it is dual to type
0A noncritical string theory only in the special case
$\mu'={\bar\mu}$. With this restriction, the space is nonsingular as
long as $\mu$ has an imaginary part. From \eref{mudefs}, this will in
turn be the case as long as the cosmological constant $\mu_M$ of the
noncritical theory is nonzero, which is natural since $\mu_M$ cuts off
the strong coupling end of the Liouville direction.  Of course from
the matrix model point of view there is still a sensible string theory
when $\mu_M=0$, but one in which the standard genus expansion of the
continuum theory does not hold, and where the role of the string
coupling is played by the inverse RR flux. The region where the RR
flux is of the same order as, or larger than, the cosmological
constant has received some discussion in the
literature\cite{Kapustin:2003hi,PandoZayas:2005tu}.

The above identification leads to the following proposed equality
between type 0A string and topological B model free energies:
\beq
\label{0AP}
\cF_{0A}(\m_M,q,\tq, R=1) =
\cF^{top,DOC}\left(\mu=\frac{g_s}{2}y, \mu'=
\frac{g_s}{2}\ybar\right)
\eeq
Using arguments analogous to those described above for the bosonic string, we also
find that the RHS perturbatively factorises:
\beq 
\label{pertfact}
\cF^{top,DOC}\left(\mu=\frac{g_s}{2}y, \mu'=\frac{g_s}{2}\ybar\right)
=\cF^{top,DC}\left(\frac{g_s}{2}y\right) +
\cF^{top,DC}\left(\frac{g_s}{2} \ybar\right)
\eeq
In principle we can now investigate the validity of the above
correspondence beyond perturbation theory. This point was considered
in Refs.\cite{Danielsson:2004ti,Danielsson:2005uz}. However, the
methods used there involve manipulation of divergent series, and we
will be able to derive all our correspondences using convergent
integral representations of the relevant special functions. 

Let us see how this works in some detail. As in
Ref.\cite{Hyun:2005fq}, we consider the open-string dual of the DOC
obtained from the Gopakumar-Vafa
correspondence\cite{Gopakumar:1998ki}.  This theory lives on a
resolved orbifolded conifold (ROC) with two $P^1$'s whose (complex)
size parameter is irrelevant in the B-model but which have
respectively $N_1,N_2$ 2-dimensional B-branes wrapped over them,
where:
\bea
\label{nvalues}
N_1 &=& \frac{y}{2}= \frac{\hq}{2} + i \mu_M\nn\\
N_2 &=& \frac{\bar y}{2}= \frac{\hq}{2} - i \mu_M
\eea
The number of branes in this correspondence is inevitably complex, and
therefore a prescription is required to complexify starting from real
integer values\footnote{However, we see that the {\it total} number of
  branes in the background $N_1 + N_2=\hq$ is real and
  integer.  This is striking, and somewhat reminiscent of fractional
  branes, though we do not have an explanation of this fact.}.

In the open string description, the partition function arises as
follows. Using Eqs.(\ref{mudefs}) and (\ref{ydef}), we find:
\bea\nn
\cF^{top,DOC}\left(\mu=\frac{g_s}{2}y, 
\mu'=\frac{g_s}{2}\ybar\right)&=& 
\cF^{top,ROC}\left(N_1= \frac{y}{2}, N_2= \frac{\ybar}{2}\right)\\
&=&
\cF^{top,RC}\left(N= \frac{y}{2}\right) + \cF^{top,RC}\left(N= \frac{\ybar}{2}\right)
\eea
where in the last step, factorisation of the Hamiltonian $H(p,x)$ has
been used.

On an ordinary resolved conifold, the free energy of $N$ D-branes is given by the
log of the matrix integral:
\beq
e^{-\cF^{top,RC}(N)}=\frac{1}{\vol(U(N)}\int dM e^{-\frac{1}{2}\tr
  M^2} =\frac{(2\pi)^{\frac{N^2}{2}}}{\vol(U(N)}
\eeq
Now we use\cite{Ooguri:2002gx}
\beq
\vol(U(N)) = \frac{(2\pi)^{\half(N^2+N)}}{G_2(N+1)}
\eeq
where $G_2(x)$ is the Barnes double-$\Gamma$ function\cite{Adamchik}
defined by:
\beq
G_2(z+1)=\Gamma(z) G_2(z),~~ G_2(1)=1
\eeq
Thus we find 
\beq
-\cF^{top,RC}\left(N= \frac{y}{2}\right) -
\cF^{top,RC}\left(N= \frac{\ybar}{2}\right) = \left(\log
G_2\left(\frac{y}{2}+1\right) -\frac{y}{4}  \log 2\pi\right) + c.c.
\eeq
Let us compare the above with what we know about the noncritical
string starting from the matrix model. In Ref.\cite{Maldacena:2005he}
the authors have given a complete nonperturbative solution for the
free energy of Type 0 noncritical strings at arbitrary radius $R$. The
free energy of type 0A theory is given by:
\beq
\label{0ANP}
-\cF_{0A}(\m_M,q,\tilde{q},R)=\Om(y,R)+\Om(\ybar,R)+\frac{\pi\m_M
R}2(|q|-|\tilde{q}|)
\eeq
where the function $\Omega$ is defined by the convergent (for ${\rm Re}\, y
>-\left(1+\frac{1}{R}\right)$) integral:
\beq
\label{omegadef}
\Omega(y,R) \equiv - \int_0^\infty
\frac{dt}{t}\left[\frac{e^{-\frac{yt}{2}}}{4 \sinh \frac{t}{2}\sinh
    \frac{t}{2R}}
-\frac{R}{t^2}+\frac{Ry}{2t}+ \Bigg(\frac{1}{24}\Big(R+\frac{1}{R}\Big)
-\frac{Ry^2}{8}\Bigg)e^{-t}\right]
\eeq
At the special radius $R=1$ it is easily shown from the integral form that:
\beq
\Omega(y,R=1) = \log G_2\left(\frac{y}{2} +1\right) - \frac{y}{4} \log 2\pi
\eeq
where $G_2$ is the Barnes function discussed above.

If we temporarily ignore the last term in \eref{0ANP}, we see that the
free energy is the sum of holomorphic and antiholomorphic
contributions. Moreover, each of these is known to be the
(complexified) free energy of the bosonic $c=1$ string at radius
$R$\cite{Klebanov:1991qa}. This is in agreement with
Eqs.(\ref{0AP}),(\ref{pertfact}).

However, the last term in \eref{0ANP} does not seem to come from the
topological string. We will discuss this issue in the following
section. First we will generalise the considerations of this
subsection to the case where the radius of the time circle is
different from $R=1$, in particular to integer radii. We will also
comment on the case of rational radii $R=\frac{p}{p'}$.

\subsection{Integer radius}

We have seen that the $c=1$ bosonic string at $R=p$ (an integer
multiple of the self-dual radius $R=1$) is dual to a topological
string living on a $Z_n$ orbifold of the conifold. An analogous result
has been proposed for the type 0A string\cite{Danielsson:2005uz}. We
will provide a simple and general derivation of this result using only
properties of convergent integral representations.

Inserting the value $R=p$ into the expression for $\Omega$,
\eref{omegadef}, we rewrite the first term in the integrand:
\beq
\frac{e^{-\frac{yt}{2}}}{4\sinh\frac{t}{2}\sinh\frac{t}{2p}}
\to 
\frac{e^{-\frac{yt}{2}}}{4(\sinh\frac{t}{2})^2}
\frac{\sinh\frac{t}{2}}{\sinh\frac{t}{2p}}
\eeq
Next, use:
\beq
\label{ratioformula}
\frac{\sinh\frac{t}{2}}{\sinh\frac{t}{2p}}
= \sum_{k=1}^p e^{\frac{t}{2p}(p-(2k-1))}
\eeq
Now define:
\beq
\label{ykdef}
y_{k} = y + \frac{-p+(2k-1)}{p}, \quad
k=1,2,\ldots,p
\eeq
Using the identities:
\bea
\label{sumidentities}
\sum_{k=1}^p \frac{1}{t^2} &=& \frac{p}{t^2}\nn\\
\sum_{k=1}^p \frac{y_{k}}{2t} &=& \frac{py}{2t}\nn\\
\sum_{k=1}^p\left(\frac{1}{12}-\frac{y_{k}^2}{8}\right)
 &=& \frac{1}{24}\left(p+ \frac{1}{p}\right) -
\frac{py^2}{8}
\eea
of which only the third one is not completely obvious, but nonetheless
easy to prove. It follows that:
\beq
\label{omegafact}
\Omega\Big(y,R=p\Big) = \sum_{k=1}^p
\Omega(y_{k},R=1) 
\eeq
We see that the free energy at rational radius factorises into $2p$
distinct contributions, of which $p$ are holomorphic in $y$ and the
remaining are anti-holomorphic. Each of the contributions corresponds
to a theory at $R=1$, or equivalently to the contribution of
topological B-branes. The factorisation is exact.

Let us analyse this in some more detail. First, by definition ${\rm
  Re}\,y\ge 0$, which not only ensures convergence of the LHS of
\eref{omegafact}, but also ensures that the RHS is convergent since this
implies that ${\rm Re}\, y_{k}>-\left(1+\frac{1}{R}\right) =-2$ for all
$k$. Therefore the equality is between convergent integral
representations as promised. 

From the above result we can conclude that for type 0 string theory at
every integer radius $R=p$, there is an exact
noncritical-topological correspondence where the corresponding
topological string lives on a $Z_{2p}$
orbifold\cite{Ghoshal:1992kx} of the conifold, whose deformed version
is:
\beq
\label{genorbconif}
zw - \prod_{k=1}^{k=p} (px-\mu_{k})
\prod_{k=1}^{k=p} (px-{\bar\mu}_{k})
\eeq
where
\beq
\label{mukdef}
\mu_{k} = \frac{g_s}{2}y_{k}
\eeq
and $y_{k}$ are defined in \eref{ykdef}. This manifold has
$2p$ independent 3-cycles that occur in complex conjugate pairs. The
factorisation into contributions from these cycles is
nonperturbatively exact upto non-universal terms, and even those terms
vanish identically at integer 0A radius.

The resolved version of this correspondence would involve the same
$Z_{2p}$ orbifold of the conifold but now with the $2p$
singularities blown up into $P^1$'s with $N_{k}$ B-branes wrapped
over each of the first $p$ cycles, and the complex conjugate number
of branes on the remaining $p$ cycles, where:
\beq
N_{k} = \frac{y_{k}}{2}
\eeq
As before, the partition function in this picture arises from the
${\rm vol}(U(N)$ factors associated to each set of $N_{k}$ branes,
giving the most direct derivation of the noncritical-topological
correspondence. 

This generalised correspondence too can be phrased in
topological-anti-topological language. In this case the topological
theory lives on a $Z_p$ orbifold, with $p$ cycles
labelled by an integer $k$ and $N_{k}$ branes wrapped
on each of them. The remaining contribution to the free energy arises
on combining with the anti-topological version of this theory.

\subsection{Rational radius}

Let us now consider more general rational radii of the form
$R=\frac{p}{p'}$, with $p$ and $p'$ co-prime.  A similar derivation to
the previous one goes through in this case, though the interpretation
presents some subtleties that we will discuss.

Inserting the value of $R$ into the expression for $\Omega$,
\eref{omegadef}, we send $t\to \frac{t}{p'}$ and then rewrite the
first term in the integrand:
\beq
\frac{e^{-\frac{yt}{2p'}}}{4\sinh\frac{t}{2p'}\sinh\frac{t}{2p}}
\to 
\frac{e^{-\frac{yt}{2p'}}}{4(\sinh\frac{t}{2})^2}
\frac{\sinh\frac{t}{2}}{\sinh\frac{t}{2p'}}
\frac{\sinh\frac{t}{2}}{\sinh\frac{t}{2p}}
\eeq
Using \eref{ratioformula} and defining:
\beq
\label{ykkprimedef}
y_{k,k'} = \frac{y-p'+(2k'-1)}{p'} + \frac{-p+(2k-1)}{p}, \quad
k=1,2,\ldots,p;\quad k'=1,2,\ldots,p' 
\eeq
we find the following identities, generalising \eref{sumidentities}:
\bea
\sum_{k=1}^p\sum_{k'=1}^{p'} \frac{1}{t^2} &=& \frac{pp'}{t^2}\nn\\
\sum_{k=1}^p\sum_{k'=1}^{p'} \frac{y_{k,k'}}{2t} &=& \frac{py}{2t}\nn\\
\sum_{k=1}^p\sum_{k'=1}^{p'}
\left(\frac{1}{12}-\frac{y_{k,k'}^2}{8}\right)
 &=& \frac{1}{24}\left(\frac{p}{p'}+ \frac{p'}{p}\right) -
\frac{py^2}{8p'}
\eea
Thus we find:
\beq
\label{omegafactrational}
\Omega\Big(y,R=\frac{p}{p'}\Big) = \sum_{k'=1}^{p'}\sum_{k=1}^p
\Omega(y_{k,k'},R=1)  -\Bigg(\frac{1}{24}\Big(\frac{p}{p'}+\frac{p'}{p}\Big) -
\frac{p y^2}{8p'}\Bigg)\log p'
\eeq
Thus, at rational radius the free energy factorises into $2pp'$
distinct contributions, of which $pp'$ are holomorphic in $y$ and the
remaining are anti-holomorphic. However, in general the factorisation
is exact only upto an analytic and therefore non-universal term.  If
we consider the special case of $p'=1$, corresponding to integer
radius in the type 0A theory, then the non-universal term vanishes. On
the other hand if we take $p=1$, corresponding to even integer radius
in the type 0B theory, then the non-universal term is
present. Subtracting the two expressions (after scaling $y\to ym$ in
one of them) we find:
\beq
\Omega\left(ym,R=\frac{1}{m}\right) - 
\Omega\left(y,R=m\right) =
-\Bigg(\frac{1}{24}\big(m+\frac{1}{m}\big) - \frac{y^2}{8m}\Bigg)\log m
\eeq
which is precisely Eq.(A.39) of \cite{Maldacena:2005he}. There, we see
that the apparent violation of T-duality by the extra term is actually
harmless and can be understand as due to the difference in natural
cutoffs for type 0A and 0B. This explains the presence of the
non-universal term, and confirms that its presence can be ignored.

We would now like to interpret the above factorisation property in
terms of contributions from singularities. For the bosonic string, the 
original ground ring analysis of Ref.\cite{Ghoshal:1992kx} tells us
that the (singular) ring at $R=\frac{p}{p'}$ is a $Z_{p}\times Z_{p'}$
orbifold of the conifold. Assuming that in type 0 strings the
parameters are complexified and occur in complex-conjugate pairs, we
expect in this case to find a $Z_{2p}\times Z_{2p'}$ orbifold of the form:
\beq
\label{pprimeorb}
\prod_{k'=1}^{p'} (zw-\alpha_{k'}) \prod_{k'=1}^{p'} (zw-{\bar\alpha}_{k'}) 
=\prod_{k=1}^{p} (px-\beta_k)\prod_{k=1}^{p} (px-{\bar\beta}_k)
\eeq
for some set of $p+p'$ complex parameters $\alpha_{k'},\beta_k$. Such
a space no longer has an interpretation as a fibration over a Riemann
surface and the analysis of its partition function is therefore more
complicated. We expect that for some (not necessarily simple)
choice of the parameters, the free energy on this space can be written
as a sum of terms as in \eref{omegafactrational} but will not be able
to show this here.

An alternate interpretation of the factorised free energy is that it
corresponds to a $Z_{2pp'}$ orbifold of the conifold:
\beq
\label{orbconifrational}
zw - \prod_{\doublet{k=1}{k'=1}}^{\doublet{k=p}{k'=p'}} (px-\mu_{k,k'})
\prod_{\doublet{k=1}{k'=1}}^{\doublet{k=p}{k'=p'}} (px-{\bar\mu}_{k,k'})
\eeq
where
\beq
\label{mukkdef}
\mu_{k,k'} = \frac{g_s}{2}y_{k,k'}
\eeq
and $y_{k,k'}$ are defined in \eref{ykkprimedef}. This manifold has
$2pp'$ independent 3-cycles that occur in complex conjugate pairs.
The resolved version of this space has the $2pp'$ singularities blown
up into $P^1$'s with $N_{k,k'}$ B-branes wrapped over each of the
first $pp'$ cycles, and the complex conjugate number of branes on the
remaining $pp'$ cycles, where:
\beq
N_{k,k'} = \frac{y_{k,k'}}{2}
\eeq
The advantage of this latter interpretation is that it preserves the
fibred structure of the manifold with a Riemann surface as the base,
and therefore all previous computations manifestly go through in the
same way. Unfortunately this interpretation is at variance with the
original proposal\cite{Ghoshal:1995wm} that the variety occurring on
the topological B-model side is in correspondence with the ground ring
on the noncritical side.

\section{Disc amplitudes and noncompact branes}

\subsection{$R=1$}

In the correspondence between noncritical type 0A strings and the
B-model on the conifold \eref{doconif} (and more generally
\eref{genorbconif}) that we have discussed above, there is right away a
puzzle. The former depends on three parameters, $q,\tq,\mu_M$, which
in the continuum Liouville description arise as the two independent RR
fluxes and the cosmological constant (in the matrix model description
these three parameters arise as a net D-brane number, a Chern-Simons
term and the Fermi level
respectively\cite{Douglas:2003up,Maldacena:2005he}). However the
topological dual only depends on the complex number $y=|q|+|\tq|+
2i\mu_M$, and therefore on only two of these three parameters. It
reproduces most of the free energy, which indeed depends only on two
parameters and is the sum of mutually complex conjugate
terms. However, the extra term in the free energy:
\beq
\label{extraterm}
{\cal F}^{disc,2} = -\frac{\pi R}{2}\mu_M (|q|-|\tq|)
\eeq
is unaccounted for (the reason for the label on this contribution will
become clear shortly). 

This term is responsible for an important effect. From the factorised
part of the free energy one extracts the following disc contribution
in the limit of large $\mu$ and fixed $\hq$\cite{Maldacena:2005he}:
\beq
\cF^{disc,1} = +\frac{\pi R}{2}|\mu_M|(|q|+|\tq|)
\eeq
Hence the total disc amplitude is:
\beq
\cF^{disc}=\cF^{disc,1}+\cF^{disc,2} = \frac{\pi R}{2}\Big[
  \big(|\mu_M|-\mu_M\big)|q| +  \big(|\mu_M|+\mu_M\big)|\tq|\Big]
\eeq
This can be written as:
\bea \cF^{disc} &=& (2\pi R)\frac{\mu_M}{2}|\tq|, \quad\mu_M>0\nn\\ &=&
(2\pi R)\frac{|\mu_M|}{2}|q|, \quad\mu_M<0 
\eea 
The physical interpretation is that for $\mu_M>0$ the RR flux of $\tq$
units associated to the gauge field $\tA$ is supported by $|\tq|$ ZZ
branes in the vacuum, with the contribution per brane to the free
energy being given by the product of its extent in Euclidean time
($2\pi R$) and its tension ($\frac{|\mu_M|}{2}$).  The other flux of
$q$ units associated to the gauge field $A$ has no source. Similarly
for $\mu_M<0$ the vacuum contains $|q|$ ZZ branes sourcing the first
flux while the other flux of $\tq$ units is not supported by any
source..

Note that in the absence of the term ${\cal F}^{disc,2}$ there is no
satisfactory physical interpretation of the disc amplitude in terms of
ZZ branes. This makes the term extremely important for a consistent
noncritical string theory. 

We now propose that the missing term is supplied, on the topological
side, by noncompact B-branes wrapping a degenerate fibre of the
Calabi-Yau over the Riemann surface $H(p,x)=0$. Such branes have been
extensively studied in Refs.\cite{Aganagic:2000gs,Aganagic:2003qj}
where they have been shown to give rise to the Kontsevich parameters
of topological matrix models. These branes are, in particular,
fermionic. Since we are considering the free energy of the string
theory, we work in the vacuum where such Kontsevich branes are
absent. However, as we now explain, it is still possible to place
noncompact branes at infinity on the Riemann surface and they can
reproduce just the desired term in the free energy.

Consider the case $R=1$. Suppose we place a single noncompact
B-brane along one branch of the degenerate fibre over a point $x$
on the Riemann surface. We would like to isolate its contribution to
the free energy compared with that of a brane at a fixed reference
position $x_*$, or in other words we assume that the brane is
asymptotically at $x_*$ but its interior region has been moved
to $x$. The action of such a brane has been
shown\cite{Aganagic:2000gs,Aganagic:2003qj} to be\footnote{In the
  language of Ref.\cite{Aganagic:2003qj}, we place the branes in the
  ``$x$-patch'' and never move them to the ``$p$-patch''.}:
\beq
\label{ncformula}
S(x) = \frac{1}{g_s}\int_{x_*}^x p(z)\,dz
\eeq

As we have seen, for the case of interest to us the Riemann surface
consists of two disjoint factors:
\beq
xp=\frac{g_s}{2}y, \quad xp=\frac{g_s}{2}\ybar
\eeq
Thus a brane on the first branch contributes:
\beq
S(x) = \frac{\mu}{g_s}\ln\frac{x}{x_*}
\eeq
Let us now place one noncompact brane above each of the two branches,
and take their asymptotic positions to be at $x_*, x'_*$ which
will both be sent to infinity. Then their total
contribution to the free energy is:
\beq
S(x,x') =  \frac{1}{2}\left(y\,\ln\frac{x}{x_{*}} + \ybar\,
\ln\frac{x'}{x'_{*}}\right) 
\eeq
Now we will choose our branes such that $x,x'$ are also at
infinity, but rotated by angles $\theta,\theta'$ respectively along
the circle at infinity relative to the original points $x_{*},
x'_{*}$. Namely:
\beq
x = x_{*}\, e^{i\theta},\quad  x' = x'_{*}\, e^{i\theta'}
\eeq
It follows that:
\bea
\label{nccontrib}
S(x_1,x_2) &=& \frac{i}{2} (y\,\theta + \ybar\,\theta')\nn\\
&=& -\mu_M(\theta-\theta') + i\frac{\hq}{2}\,(\theta + \theta')
\eea
The factors of $g_s$ have conveniently cancelled out, and the real part of
the above contribution is proportional to $\mu_M$. Now if we choose:
\beq
\label{thetaangles}
\theta = -\theta' = \frac{\pi}{4} (|q|-|\tq|)
\eeq
we find that the noncompact branes give a contribution:
\beq
S = -\frac{\pi}{2}\mu_M(|q|-|\tq|)
\eeq
to the free energy, precisely equal to that in \eref{extraterm} at $R=1$.

To summarise, we have shown that if we place a noncompact B-brane at
$x\to\infty$ on each branch of the Riemann surface
\beq
H(p,x)=(px-\mu)(px-\mu')=0
\eeq
and moreover require that the branes wind at infinity by the angles in
\eref{thetaangles}, we precisely reproduce the disc contribution to the
free energy of \eref{extraterm}. This situation is depicted in
Fig.\ref{riemann}.

\FIGURE{
\label{riemann}
\epsfxsize=11cm
\epsfbox{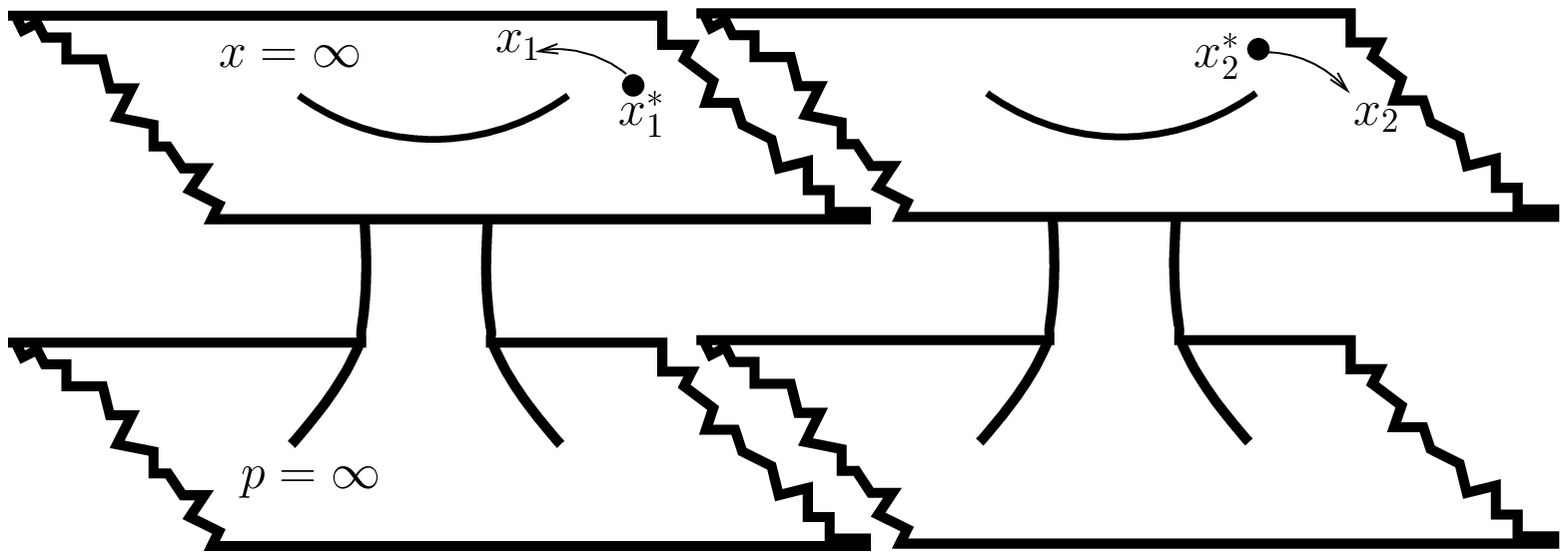}
\caption{The Riemann surface with noncompact branes at infinity.}}

This then completes the definition of
the topological dual to type 0A strings at the special radius.

The above system also has a description in
topological-anti-topological language. As we have seen, the
topological theory then lives on the pure conifold, having a Riemann
surface with only one branch. Now we place a single noncompact brane
on it with winding angle $\theta$ given by \eref{thetaangles}. Adding
the anti-topological theory introduces the second noncompact brane
with winding $-\theta$ and we recover the correct free energy.

\subsection{Integer and rational radius}

Let us now extend these considerations to other integer radii. At
radius $R=p$, we have the possibility of placing noncompact branes at
infinity on each of $2n$ branches of the Riemann surface $H(x,p)=0$
obtained from \eref{genorbconif}. Parametrising the angles by which
these branes wind as:
\beq
x_{i} = x_{*i}\,e^{i\,\theta_i},\quad x'_{i} = x'_{*i}\,e^{i\,\theta'_i}, 
\eeq
the contribution of these branes to the free energy is:
\bea
S(x_i, x'_i) &=& \frac{i}{2}\sum_{j=1}^n \left(y_j \,\theta_j + \ybar_j\,
\theta'_j\right)\nn\\ 
&=& -\mu_M\sum_{j=1}^n (\theta_j -\theta'_j) + 
i\sum_{j=1}^n \frac{\hq_j}{2}(\theta_j +\theta'_j)
\eea
where
\beq
\hq_j = \hq -1 + \frac{2j-1}{n},\quad j=1,2,\ldots,n
\eeq
It is natural to take
\beq
\theta_j = -\theta'_j = \frac{\pi}{4}\left(|q|-|\tq|\right),\quad \hbox{all}~ j=1,2,\ldots,n
\eeq
which leads to a contribution to the free energy:
\beq
S = -\frac{\pi n}{2}\mu_M(|q|-|\tq|)
\eeq
in precise agreement with \eref{extraterm} for $R=n$.

It appears as if in this case the noncompact brane configuration is
not unique. However, note that choosing $\theta_j=-\theta'_j$ for all
$j$ is essential to make the free energy real. After this, the choice
we have made is the most symmetric one which gives the correct disc
amplitude.

In the topological-anti-topological approach, we would instead have
$p$ branches in the Riemann surface and therefore $p$ noncompact
branes with associated angles $\theta_{k}$. The remaining
noncompact branes with angles $-\theta_{k}$ then arise on the
anti-topological side.

It is quite nontrivial that we were able to reproduce the subtle disc
term by a simple configuration of noncompact branes in every case. The
scaling with $g_s$ of the holomorphic Chern-Simons action and of the
complex-structure moduli $\mu_{k,k'}$ defined in \eref{mukkdef}
exactly cancel out. Moreover, $\mu_{k,k'}$ all have a common imaginary
part proportional to $\mu_M$. These facts were important in allowing
us to obtain the desired contribution from noncompact branes.

Now let us briefly consider rational radius.  If we accept the $Z_{2p}\times
Z_{2p'}$ orbifold interpretation of \eref{pprimeorb} then it is not clear how
to extend the above considerations to radius $R=\frac{p}{p'}$. This is
because the manifold is no longer of the form $zw=H(p,x)$ and
therefore the Riemann surface interpretation itself needs to be
generalised, which lies beyond the scope of the present work.

\section{Discussion}

One of our main results has been that the noncritical-topological
correspondence for type 0 noncritical strings has to include
noncompact branes on the topological side. This introduces a
dependence on a new parameter which we interpret as $|q|-|\tq|$ on the
noncritical side, and renders the duality consistent with the
dependence of the noncritical theory on three parameters: $\mu_M, q $
and $\tq$.

The identification between the phases of noncompact branes and the
parameter $|q|-|\tq|$, via \eref{thetaangles}, appears rather ad
hoc. From \eref{thetaangles} it is tempting to imagine that there
could be a missing normalisation factor of 8 which changes
$\frac{\pi}{4}$ to $2\pi$. In that case one could have postulated that
the noncompact branes have an integer winding at infinity and this
integer gets identified with the integer $|q|-|\tq|$. This would make
the identification a little less ad hoc. However we did not find such
a missing normalisation factor.

Given that the subtle disc term is required in the noncritical string
by consistency, one may ask if the presence of noncompact branes in
the topological theory is also a consistency requirement. However,
this seems not to be the case. On the noncritical side there is the
possibility of ZZ branes in the vacuum, and it is only after including
the subtle term that the vacuum has a definite intepretation as
containing or not containing such branes. However ZZ branes do not (so
far) have a direct analogue on the topological side and so it is
possible that the topological theory without the subtle disc term, and
hence with an exactly holomorphically factorised free energy, is
consistent by itself. The only thing that would fail is its
correspondence to the noncritical theory. Nevertheless it would be
interesting if there were a way to understand ZZ branes from the
noncritical side. It would be equally interesting to understand the
presence\cite{Maldacena:2005he} of $q\tq$ fundamental strings in the
vacuum, for which we have not found a direct topological explanation.

We also found that the free energy of the full type 0A theory has a
nonperturbatively exact factorisation into contributions from compact
and noncompact branes. Apparently there is no room for any
interactions between these different branes, or in other words the
open strings stretched among any two of these branes (both compact, or
both noncompact, or one of each) seem to decouple completely. This is
somewhat puzzling but must be related in some way to the topological
nature of the theory as well as to having distinct branches of the Riemann
surface $H(p,x)=0$.

It is amusing that compact and noncompact branes make use of different
pieces of the holomorphic Chern-Simons theory restricted to a
2-cycle\cite{Dijkgraaf:2002fc}:
\beq
S = \frac{1}{g_s}\int \tr(\Phi_1{\bar D}\Phi_0 + W(\Phi_0)\omega)
\eeq
For compact branes, the first term can be shown to be irrelevant while
the second one gives a matrix-valued superpotential, which for our
case is simply an independent quadratic for each branch of the Riemann
surface. For noncompact branes it is the second term which is
irrelevant (because the volume is infinite, we subtract the free
energy of deformed compact branes from the undeformed
ones\cite{Aganagic:2000gs}) while the first term leads to the
expression $\int p dx$. In this case there is no matrix model, because
we have placed only one brane on each branch.

It is clearly of interest to generalise our construction to include
more noncompact branes that act as sources for incoming closed-string
tachyons on the noncritical side\footnote{Our noncompact branes do not
  act as such sources precisely because they are located at
  $x\to\infty$.}, as well as non-normalisable deformations of the
  conifold which are associated to outgoing
  tachyons\cite{Aganagic:2003qj}. When carried out for the general
  orbifolded conifold \eref{genorbconif}, this will provide the
  analogue of the Normal Matrix
  Model\cite{Alexandrov:2003qk,Mukherjee:2005aq} for type 0 strings,
  valid for all rational radius. This is an important generalisation
  of the KP\cite{Imbimbo:1995yv} model which has already been found
  using the topological string construction for both bosonic $c=1$
  strings\cite{Aganagic:2003qj} and type 0 strings\cite{Ita:2004yn}.

As is well known, topological descriptions of noncritical strings are
simplest at $R=1$ in appropriate units, and can then be generalised to
integer multiples of this radius, as done before for bosonic strings
and in this paper for type 0 strings. In this way we can describe the
Euclidean or finite-temperature version of the theory. To get to the
zero-temperature case one then has to take the limit
$R\to\infty$. This limit has been explored before, most recently in
Ref.\cite{Dijkgraaf:2003xk} where it was related to
deconstruction. Our analysis in the present work can potentially add
something to this story. Consider \eref{ykkprimedef} at $p'=1$ and
take $p\to\infty$.  In this limit we find that $y_{k,k'}$ varies
continuously in the open interval $(y-1,y+1)$. From \eref{ydef} this
amounts to saying that the RR flux effectively varies continuously
over the same interval. This suggests a higher-dimensional origin and
may again link the topological theory to deconstruction in some way.

As we have seen, all our considerations extend to the
topological-anti-topological picture of Ref.\cite{Danielsson:2005uz},
which seems more natural in one sense. The $Z_{2p}$
orbifolded conifold has in principle $2p$ independent complex
structure parameters $\mu_{k},{\bar\mu}_{k}$. The
noncritical-topological correspondence requires half of them to be
constrained to be complex conjugates of the other half, which is
naturally achieved if we think of the system in the
topological-anti-topological way. In that case only the $p$
parameters $\mu_{k}$ can be independent. However, as we have seen,
the $\mu_{k}$ are all determined in terms of two parameters
embodied in $y$, and the topological-anti-topological picture does not
seem to help in explaining this fact. Therefore, if it is to be
genuinely useful, perhaps it needs to be extended to a generalised
principle where the holomorphic part of the free energy further
factorises into contributions from $p$ independent theories.

\section*{Acknowledgements}

We would like to thank Debashis Ghoshal, Rajesh Gopakumar, K. Narayan
and Ashoke Sen for very helpful discussions, and Sang-Heon Yi for a
useful correspondence. Part of this work was carried out in the
stimulating and pleasant environment of the Harish-Chandra Institute,
whose hospitality we gratefully acknowledge. We thank the people of
India for generously supporting our research. The work of A.M. was
supported in part by CSIR Award No. 9/9/256(SPM-5)/2K2/EMR-I.

\bibliographystyle{JHEP}

\bibliography{discterm}

\providecommand{\href}[2]{#2}\begingroup\raggedright\begin{thebibliography}{10}

\bibitem{Witten:1991mk}
E.~Witten, {\it The {N} matrix model and gauged {WZW} models},  {\em Nucl.
  Phys.} {\bf B371} (1992) 191--245.

\bibitem{Mukhi:1993zb}
S.~Mukhi and C.~Vafa, {\it Two-dimensional black hole as a topological coset
  model of $c = 1$ string theory},  {\em Nucl. Phys.} {\bf B407} (1993)
  667--705, [\href{http://xxx.lanl.gov/abs/hep-th/9301083}{{\tt
  hep-th/9301083}}].

\bibitem{Ghoshal:1993qt}
D.~Ghoshal and S.~Mukhi, {\it Topological {L}andau-{G}inzburg model of
  two-dimensional string theory},  {\em Nucl. Phys.} {\bf B425} (1994)
  173--190, [\href{http://xxx.lanl.gov/abs/hep-th/9312189}{{\tt
  hep-th/9312189}}].

\bibitem{Hanany:1994fi}
A.~Hanany, Y.~Oz, and M.~Ronen~Plesser, {\it Topological {L}andau-{G}inzburg
  formulation and integrable structure of 2-{D} string theory},  {\em Nucl.
  Phys.} {\bf B425} (1994) 150--172,
  [\href{http://xxx.lanl.gov/abs/hep-th/9401030}{{\tt hep-th/9401030}}].

\bibitem{Ghoshal:1994rs}
D.~Ghoshal, C.~Imbimbo, and S.~Mukhi, {\it Topological 2-{D} string theory:
  Higher genus amplitudes and ${W}_\infty$ identities},  {\em Nucl. Phys.} {\bf
  B440} (1995) 355--372, [\href{http://xxx.lanl.gov/abs/hep-th/9410034}{{\tt
  hep-th/9410034}}].

\bibitem{Ghoshal:1995wm}
D.~Ghoshal and C.~Vafa, {\it $c = 1$ string as the topological theory of the
  conifold},  {\em Nucl. Phys.} {\bf B453} (1995) 121--128,
  [\href{http://xxx.lanl.gov/abs/hep-th/9506122}{{\tt hep-th/9506122}}].

\bibitem{Witten:1991zd}
E.~Witten, {\it Ground ring of two-dimensional string theory},  {\em Nucl.
  Phys.} {\bf B373} (1992) 187--213,
  [\href{http://xxx.lanl.gov/abs/hep-th/9108004}{{\tt hep-th/9108004}}].

\bibitem{Ita:2004yn}
H.~Ita, H.~Nieder, Y.~Oz, and T.~Sakai, {\it Topological {B}-model, matrix
  models, ${\hat c} = 1$ strings and quiver gauge theories},  {\em JHEP} {\bf
  05} (2004) 058, [\href{http://xxx.lanl.gov/abs/hep-th/0403256}{{\tt
  hep-th/0403256}}].

\bibitem{Danielsson:2004ti}
U.~H. Danielsson, M.~E. Olsson, and M.~Vonk, {\it Matrix models, 4d black holes
  and topological strings on non-compact calabi-yau manifolds},  {\em JHEP}
  {\bf 11} (2004) 007, [\href{http://xxx.lanl.gov/abs/hep-th/0410141}{{\tt
  hep-th/0410141}}].

\bibitem{Hyun:2005fq}
S.~Hyun, K.~Oh, J.-D. Park, and S.-H. Yi, {\it Topological {B}-model and ${\hat
  c} = 1$ string theory},  {\em Nucl. Phys.} {\bf B729} (2005) 135--162,
  [\href{http://xxx.lanl.gov/abs/hep-th/0502075}{{\tt hep-th/0502075}}].

\bibitem{Danielsson:2005uz}
U.~H. Danielsson, N.~Johansson, M.~Larfors, M.~E. Olsson, and M.~Vonk, {\it 4d
  black holes and holomorphic factorization of the 0a matrix model},  {\em
  JHEP} {\bf 10} (2005) 046,
  [\href{http://xxx.lanl.gov/abs/hep-th/0506219}{{\tt hep-th/0506219}}].

\bibitem{Douglas:2003up}
M.~R. Douglas {\em et~al.}, {\it A new hat for the $c = 1$ matrix model},
  \href{http://xxx.lanl.gov/abs/hep-th/0307195}{{\tt hep-th/0307195}}.

\bibitem{Maldacena:2005he}
J.~M. Maldacena and N.~Seiberg, {\it Flux-vacua in two dimensional string
  theory},  {\em JHEP} {\bf 09} (2005) 077,
  [\href{http://xxx.lanl.gov/abs/hep-th/0506141}{{\tt hep-th/0506141}}].

\bibitem{Mukhi:2003sz}
S.~Mukhi, {\it Topological matrix models, {L}iouville matrix model and $c = 1$
  string theory},  \href{http://xxx.lanl.gov/abs/hep-th/0310287}{{\tt
  hep-th/0310287}}.

\bibitem{Antoniadis:1995zn}
I.~Antoniadis, E.~Gava, K.~S. Narain, and T.~R. Taylor, {\it ${N}=2$ type {II}
  heterotic duality and higher derivative {F} terms},  {\em Nucl. Phys.} {\bf
  B455} (1995) 109--130, [\href{http://xxx.lanl.gov/abs/hep-th/9507115}{{\tt
  hep-th/9507115}}].

\bibitem{Morales:1997zv}
J.~F. Morales and M.~Serone, {\it {BPS} states and supersymmetric index in ${N}
  = 2$ type {I} string vacua},  {\em Nucl. Phys.} {\bf B501} (1997) 427--444,
  [\href{http://xxx.lanl.gov/abs/hep-th/9703049}{{\tt hep-th/9703049}}].

\bibitem{Gukov:2005iy}
S.~Gukov, K.~Saraikin, and C.~Vafa, {\it A stringy wave function for an ${S}^3$
  cosmology},  {\em Phys. Rev.} {\bf D73} (2006) 066009,
  [\href{http://xxx.lanl.gov/abs/hep-th/0505204}{{\tt hep-th/0505204}}].

\bibitem{Ghoshal:1992kx}
D.~Ghoshal, D.~P. Jatkar, and S.~Mukhi, {\it Kleinian singularities and the
  ground ring of $c=1$ string theory},  {\em Nucl. Phys.} {\bf B395} (1993)
  144--166, [\href{http://xxx.lanl.gov/abs/hep-th/9206080}{{\tt
  hep-th/9206080}}].

\bibitem{Ghoshal:1993ur}
D.~Ghoshal, P.~Lakdawala, and S.~Mukhi, {\it Perturbation of the ground
  varieties of $c = 1$ string theory},  {\em Mod. Phys. Lett.} {\bf A8} (1993)
  3187--3200, [\href{http://xxx.lanl.gov/abs/hep-th/9308062}{{\tt
  hep-th/9308062}}].

\bibitem{Gopakumar:1998vy}
R.~Gopakumar and C.~Vafa, {\it Topological gravity as large $n$ topological
  gauge theory},  {\em Adv. Theor. Math. Phys.} {\bf 2} (1998) 413--442,
  [\href{http://xxx.lanl.gov/abs/hep-th/9802016}{{\tt hep-th/9802016}}].

\bibitem{Aganagic:2003qj}
M.~Aganagic, R.~Dijkgraaf, A.~Klemm, M.~Marino, and C.~Vafa, {\it Topological
  strings and integrable hierarchies},  {\em Commun. Math. Phys.} {\bf 261}
  (2006) 451--516, [\href{http://xxx.lanl.gov/abs/hep-th/0312085}{{\tt
  hep-th/0312085}}].

\bibitem{Gukov:2003yp}
S.~Gukov, T.~Takayanagi, and N.~Toumbas, {\it Flux backgrounds in 2d string
  theory},  {\em JHEP} {\bf 03} (2004) 017,
  [\href{http://xxx.lanl.gov/abs/hep-th/0312208}{{\tt hep-th/0312208}}].

\bibitem{Ooguri:2004zv}
H.~Ooguri, A.~Strominger, and C.~Vafa, {\it Black hole attractors and the
  topological string},  {\em Phys. Rev.} {\bf D70} (2004) 106007,
  [\href{http://xxx.lanl.gov/abs/hep-th/0405146}{{\tt hep-th/0405146}}].

\bibitem{Ooguri:2005vr}
H.~Ooguri, C.~Vafa, and E.~P. Verlinde, {\it Hartle-hawking wave-function for
  flux compactifications},  {\em Lett. Math. Phys.} {\bf 74} (2005) 311--342,
  [\href{http://xxx.lanl.gov/abs/hep-th/0502211}{{\tt hep-th/0502211}}].

\bibitem{Kapustin:2003hi}
A.~Kapustin, {\it Noncritical superstrings in a {R}amond-{R}amond background},
  {\em JHEP} {\bf 06} (2004) 024,
  [\href{http://xxx.lanl.gov/abs/hep-th/0308119}{{\tt hep-th/0308119}}].

\bibitem{PandoZayas:2005tu}
L.~A. Pando~Zayas and D.~Vaman, {\it Condensing momentum modes in $2d$ 0{A}
  string theory with flux},  \href{http://xxx.lanl.gov/abs/hep-th/0507061}{{\tt
  hep-th/0507061}}.

\bibitem{Gopakumar:1998ki}
R.~Gopakumar and C.~Vafa, {\it On the gauge theory/geometry correspondence},
  {\em Adv. Theor. Math. Phys.} {\bf 3} (1999) 1415--1443,
  [\href{http://xxx.lanl.gov/abs/hep-th/9811131}{{\tt hep-th/9811131}}].

\bibitem{Ooguri:2002gx}
H.~Ooguri and C.~Vafa, {\it Worldsheet derivation of a large-{N} duality},
  {\em Nucl. Phys.} {\bf B641} (2002) 3--34,
  [\href{http://xxx.lanl.gov/abs/hep-th/0205297}{{\tt hep-th/0205297}}].

\bibitem{Adamchik}
V.~Adamchik, {\it Contributions to the theory of the {B}arnes function},
  \href{http://xxx.lanl.gov/abs/math.CA/0308086}{{\tt math.CA/0308086}}.

\bibitem{Klebanov:1991qa}
I.~R. Klebanov, {\it String theory in two dimensions},
  \href{http://xxx.lanl.gov/abs/hep-th/9108019}{{\tt hep-th/9108019}}.

\bibitem{Aganagic:2000gs}
M.~Aganagic and C.~Vafa, {\it Mirror symmetry, {D}-branes and counting
  holomorphic discs},  \href{http://xxx.lanl.gov/abs/hep-th/0012041}{{\tt
  hep-th/0012041}}.

\bibitem{Dijkgraaf:2002fc}
R.~Dijkgraaf and C.~Vafa, {\it Matrix models, topological strings, and
  supersymmetric gauge theories},  {\em Nucl. Phys.} {\bf B644} (2002) 3--20,
  [\href{http://xxx.lanl.gov/abs/hep-th/0206255}{{\tt hep-th/0206255}}].

\bibitem{Alexandrov:2003qk}
S.~Y. Alexandrov, V.~A. Kazakov, and I.~K. Kostov, {\it 2d string theory as
  normal matrix model},  {\em Nucl. Phys.} {\bf B667} (2003) 90--110,
  [\href{http://xxx.lanl.gov/abs/hep-th/0302106}{{\tt hep-th/0302106}}].

\bibitem{Mukherjee:2005aq}
A.~Mukherjee and S.~Mukhi, {\it $c = 1$ matrix models: {E}quivalences and
  open-closed string duality},  {\em JHEP} {\bf 10} (2005) 099,
  [\href{http://xxx.lanl.gov/abs/hep-th/0505180}{{\tt hep-th/0505180}}].

\bibitem{Imbimbo:1995yv}
C.~Imbimbo and S.~Mukhi, {\it The topological matrix model of $c = 1$ string},
  {\em Nucl. Phys.} {\bf B449} (1995) 553--568,
  [\href{http://xxx.lanl.gov/abs/hep-th/9505127}{{\tt hep-th/9505127}}].

\bibitem{Dijkgraaf:2003xk}
R.~Dijkgraaf and C.~Vafa, {\it N = 1 supersymmetry, deconstruction, and bosonic
  gauge theories},  \href{http://xxx.lanl.gov/abs/hep-th/0302011}{{\tt
  hep-th/0302011}}.

\end{thebibliography}\endgroup

\end{document}